\documentclass[10pt,pra,superscriptaddress,showpacs, twocolumn, floatfix]{revtex4}
\usepackage{graphics}
\usepackage{amsmath}
\usepackage{color}
\usepackage{amsfonts}
\usepackage{bbold}
\DeclareMathOperator{\Tr}{Tr}

\renewcommand{\vec}[1]{\overset{\rightarrow}{#1}}
\usepackage {tikz}
\usetikzlibrary {positioning}
\newcommand{\braket}[2]{\left< #1 \vphantom{#2} \right|
 \left. #2 \vphantom{#1} \right>} 
\newcommand{\cev}[1]{\reflectbox{\ensuremath{\vec{\reflectbox{\ensuremath{#1}}}}}}
\newcommand{\lefta}[1]{\overleftarrow{#1}}
\newcommand{\righta}[1]{\overrightarrow{#1}}
\newcommand{\e}{\epsilon}

\newcommand{\systemname}{thermalizing qubit cloud }

\newcommand{\chinagrant}{the National Basic Research Program of China Grant 2011CBA00300, 2011CBA00302, the National Natural Science Foundation of China Grant 61033001, 61361136003}

\newtheorem{theorem}{Theorem}[section]

\newtheorem{definition}[theorem]{Definition}
\input{Qcircuit}

\begin{document}
\title{Towards Quantifying Complexity with Quantum Mechanics}

\author{Ryan Tan}
\affiliation{Center for Quantum Information, Institute for Interdisciplinary Information Sciences, Tsinghua University, 100084 Beijing, China} \affiliation{Centre for Quantum Technologies,
National University of Singapore, 3 Science Drive 2, 117543 Singapore, Singapore}

\author{Daniel R. Terno}
\affiliation{Department of Physics and Astronomy, Macquarie University, Sydney, New South Wales 2109, Australia}

\author{Jayne Thompson}
\affiliation{Centre for Quantum Technologies,
National University of Singapore, 3 Science Drive 2, 117543 Singapore, Singapore}
\email{thompson.jayne2@gmail.com}
\author{Vlatko Vedral}
\affiliation{ Department of Physics, University of Oxford, Clarendon Laboratory, Oxford, OX1 3PU, United Kingdom }
\affiliation{Centre for Quantum Technologies,
National University of Singapore, 3 Science Drive 2, 117543 Singapore, Singapore}\affiliation{Department of Physics, National University of Singapore, 2 Science Drive 3, 117551 Singapore, Singapore}
\author{ Mile Gu}
\affiliation{Center for Quantum Information, Institute for Interdisciplinary Information Sciences, Tsinghua University, 100084 Beijing, China} \affiliation{Centre for Quantum Technologies,
National University of Singapore, 3 Science Drive 2, 117543 Singapore, Singapore}

\begin{abstract}While we have intuitive notions of structure and complexity, the formalization of this intuition is non-trivial. The statistical complexity is a popular candidate. It is based on the idea that the complexity of a process can be quantified by the complexity of its simplest mathematical model - the model that requires the least past information for optimal future prediction. Here we review how such models, known as $\epsilon$-machines can be further simplified through quantum logic, and explore the resulting consequences for understanding complexity. In particular, we propose a new measure of complexity based on quantum $\epsilon$-machines. We apply this to a simple system undergoing constant thermalization. The resulting quantum measure of complexity aligns more closely with our intuition of how complexity should behave. \end{abstract}
\maketitle

Are there any universal laws governing the evolution of complexity? While the second law of thermodynamics indicates ever increasing entropy, complexity seems to behave differently. The hot, smooth plasma near the Universe's birth and the final state of thermal equilibrium predicted by its heat death both appeal to our intuition of simplicity. Yet between these extremes, where there are stars, galaxies and life, the universe is complex; and because of that it is interesting. To answer this question, one must first quantify complexity.

This task is non-trivial. While we have intuitive notions of what is interesting or complex, they are deceptively difficult to formalize. The first attempts to quantify complexity came in the form of Kolmogorov complexity \cite{kolmogorov1998tables}. This measure equates the complexity of a sequence of numbers to the size of the minimal computer program that generates the sequence. While Kolmogorov complexity correctly identifies a constant sequence, consisting entirely of 0s, as simple; it is maximized by sequences that are completely random. This makes the measure unsatisfying for characterizing complexity, because it seems to misconstrue randomness with structure~\cite{ladyman2013complex}.

A promising way to avoid this problem came from the study of computational mechanics, where one is concerned in building $\epsilon$-machines, the simplest predictive model of a supplied stochastic process~\cite{crutchfield1989inferring}. One reason is that if a stochastic process is more complex, then replicating its future statistics will require more information about its past. A completely random process, for example, requires no past information to reproduce its future statistics; and nor would a process that outputs only zero. Meanwhile, a process with less trivial behavior, such as one which alternates between zero and one on successive emissions, can only be faithfully replicated by storing its last emission - and is thus more complex. The minimum amount of past information required to optimally predict a given process thus introduces a more suitable measure of complexity. Known as the \emph{statistical complexity}, $C_\mu$, its clear operational significance and relative ease of evaluation have resulted in its widespread adoption in diverse contexts \newline \cite{shalizi2001computational,li2008multiscale,crutchfield1997statistical,wiesner2012information,Cerbus,crutchfield2012between}.

Nevertheless, statistical complexity still displays certain incongruities. Notably, it is not continuous -- infinitesimal perturbations in the statistics of a process can lead to large jumps in its statistical complexity. A physical process that asymptotically approaches total randomness can have monotonically increasing statistical complexity, even if its final steady state has statistical complexity zero \cite{crutchfield1997statistical}. This seems to contradict our intuition of what complexity should be.

The results discussed so far however, have been limited to classical logic. Reality is ultimately quantum mechanical. If quantum logic allows us as to build simpler predictive models - models that use less past information - then the quantum analogue of statistical complexity could be a more fitting quantifier of structure. Indeed, it has been recently demonstrated that quantum models can optimally predict the future statistics of a classical stochastic process while generically storing less information about the past than their classical counterparts \cite{gu2012quantum}. Systems that are complex to predict classically may be simpler quantum mechanically.

The objective of this article is to explore how quantum logic can improve our understanding of what makes a process interesting and complex, and ultimately contribute to discovering how complexity evolves. We review $\epsilon$-machines - the provably simplest classical models; and how they can be further simplified through quantum logic. This motivates us to introduce a new measure of complexity, $C_q$, based on the complexity of quantum $\epsilon$-machines.

We apply these ideas to a toy system featuring monotonically increasing entropy. First we show that the system's complexity, when characterized by $C_q$, aligns more closely with our intuition of how complexity should behave. In addition to being zero when the system features zero or maximal entropy, it is also continuous; an infinitesimal perturbation in the statistics of the process will cause an infinitesimal change in $C_q$. Thus, we highlight the relevance of quantum mechanics in studying how structure and complexity persist and evolve.

The article is structured as follows. Section \ref{sect:background} will review  statistical complexity, $\epsilon$-machines, and their extension to quantum $\epsilon$-machines. Section \ref{sect:quantcomplexity} reviews a new measure of complexity based on the complexity of these quantum $e$-machines \cite{gu2012quantum}. Subsection \ref{sect:models} then applies this measure to a toy system with monotonically increasing noise; and highlights how the complexity of quantum and classical $\epsilon$-machines diverge. Concluding remarks are presented in Section \ref{sect:discussion}.

\section{Preliminaries}\label{sect:background}

This section provides a background mechanism of inferring the statistical complexity of observed phenomena and their quantum extensions. For a more extensive treatment of these topics, see Refs. \cite{shalizi2001computational,gu2012quantum,crutchfield1994calculi}. Familiarity with quantum information to the level of \cite{nielsen2010quantum} is assumed.

\subsection{Computational Mechanics, Complexity and Predictive Models}

Computational mechanics seeks to study the complexity of systems through the lens of predictive models. The general approach is to assume that a system's behavior can probed at discrete points in time $t$, with outcomes $x_{t}\in\Sigma$ dictated by random variable $\mathbf{X}_{t}$. Here, $\Sigma$ defines the set of possible observable outcomes.

In the ideal scenario, the system may be probed indefinitely. The observable behaviour of such a system is a sequence of output values $\overset{\leftrightarrow}{x}=\cdots x_{-2}x_{-1}x_{0}x_{1}x_{2}\cdots$, where $\cev{x}=\cdots x_{-2}x_{-1}$ and $\vec{x}=x_{0}x_{1}x_{2}\cdots$ are the output sequences of the past and future respectively. This results in a stochastic process, defined by the joint probability distribution $P(\lefta{X},\righta{X})$. Here, $\lefta{X}$ and $\righta{X}$ are the random variables governing $\cev{x}$ and $\vec{x}$. Each realization of the system has a particular past $\cev{x}$ with probability $P(\lefta{X}=\cev{x})$ and exhibits a particular future $\vec{x}$ with probability $P(\righta{X}=\vec{x}|\lefta{X}=\cev{x})$.

Computational mechanics aims to infer the complexity of the system through these statistics. It asks, if we are to build a mathematical model of the process with statistically indistinguishable behavior what is the minimal amount of information it needs to keep about past observations? The more memory required the greater its complexity.

To formalize this intuition, envision our system encased in a black box, that simply outputs the outcome $x_t \in \Sigma$ at time $t \in \mathbb{Z}$. In a second black box, a computer attempts to simulate the process through execution of an appropriate mathematical model. It takes as input some $S$ that is a function of past observations, $\lefta{x}$ and outputs appropriate future statistics. For this simulation to be completely faithful the two boxes must be indistinguishable. For each instance of the process with past $\lefta{x}$, the model must output statistical predictions $\righta{x}$, according to the statistical distribution
\begin{equation}\label{eq:futureprobdistrib}
P(\overrightarrow{X}|\overleftarrow{X}=\cev{x}).
\end{equation}
The amount of information such a model requires to track is then determined by the minimal amount of space it needs to store about $\lefta{x}$. Formally this is given by the information entropy of $\mathbf{S}$, the random variable that governs input variable $S$. We refer to this as the complexity of a given predictive model.

\subsection{Classical Statistical Complexity}\label{subsect:classstatcomplex}

The statistical complexity for a process is determined by the complexity of its simplest model -- that correctly simulates the process while requiring the least amount of information about $\lefta{x}$. Thus we must find the best way of compressing the past without losing information about the future.

An immediate brute-force attempt is to store the entire past. Such a model takes the input $\cev{x}$ directly and outputs the future according to \eqref{eq:futureprobdistrib} and thus stores the input with information content $C=H(\lefta{X})$, where $H$ denotes the Shannon entropy. This is clearly not efficient. For example, for a series of fair coin tosses $P(\lefta{X},\righta{X})$ is uniform over the distribution of binary strings. Thus $C$ is infinite, implying that a simulation using this approach will require an infinite amount of memory. Clearly better approaches exist.

The simplest classical predictive models are epsilon machines ($\e$-machines) \cite{crutchfield1989inferring,shalizi2001computational}. Jointly proposed by Crutchfield and Shalizi, $\e$-machines are based on the reasoning that two different pasts need only be distinguished if they have differing future statistics. This motivates an equivalence relation on the set of pasts, $\sim$, such that for any two distinct pasts, $\cev{x}$ and $\cev{x}'$, $\cev{x} \sim\cev{x}' \iff P(\righta{X}|\lefta{X}=\cev{x})=P(\righta{X}|\lefta{X}=\cev{x}')$. Each equivalence class is referred to as a causal state $S_i$ governed by a random variable $\mathbf{S}$. Let $\mathcal{S} = \{S_{i}\}_{i=1,2,\cdots,N}$ denote the set of all causal states where $N$ represents the total number of causal states. An $\epsilon$-machine does not store $\lefta{x}$, but only stores the equivalence class to which $\lefta{x}$ belongs. More formally:

\begin{definition}[$\epsilon$-machines] \label{def:epsilonmachines} Given a stochastic process  $P(\lefta{X},\righta{X})$, we can define its $\epsilon$-machine as follows: The $\e$-machine of the process is the ordered pair $\{\e,\mathbf{T}\}$, where $\e$ is the causal state function such that $\e(\cev{x})=S_{i}$, $S_{i}\in\mathcal{S}$, $\mathbf{T}=\{T^{(r)}_{j,k}|r\in\Sigma, S_{j},S_{k}\in \mathcal{S}\}$ is a collection of transition probabilities, with $T^{(r)}_{j,k}=P(\mathbf{S}_{t}=S_{k},\mathbf{X}_{t}=r|\mathbf{S}_{t-1}=S_{j})$.
\end{definition}

After initiating an $\epsilon$-machine in state $\e(\cev{x})$, there exists standard algorithms to systematically generate desired future statistics. At each time-step $t$, an $\epsilon$-machine in causal state $S_{j}$ will emit output $r\in\Sigma$ and transit to causal state $S_{k}$, with probability $T^{(r)}_{j,k}$.

Since there are no simpler classical models, Crutchfield and Shalizi defined the statistical complexity of a given process to be synonymous with the internal entropy of the $\e$-machine, i.e.,
\begin{equation}\label{eq:cstatcomplex}
C_{\mu}=H(\mathbf{S})=-\sum_{i=1}^{N}p_{i}\log_2 p_{i}
\end{equation}
where $p_{i}=P(\mathbf{S}=S_{i}=\e(\cev{x}))$ is the probability that $\lefta{x} \in S_i$.

Statistical complexity does not equate randomness with structure. For a completely random process the conditional future for all pasts is the same. Thus only one causal state is needed to specify the process and $C_{\mu}=0$. This implies that completely random processes have no inherent structure.  On the other hand, a process that emits a constant bit-string also features zero complexity as any probability distribution of a constant variable has zero entropy. In between these extremes, statistical complexity can be expected to peak.

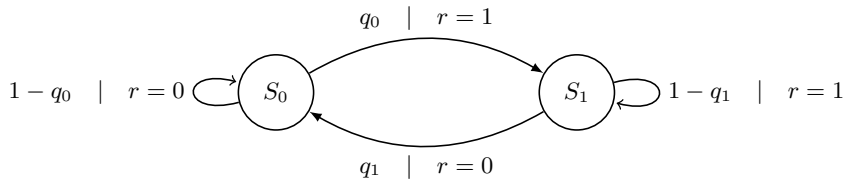
\begin{figure*}
\centering
\begin{tikzpicture}[-latex ,auto ,node distance =4 cm ,on grid ,
semithick ,
state/.style ={ circle ,top color =white ,
draw, minimum width =1 cm}]
\node[state] (C) {$S_{1}$};
\node[state] (A) [left=of C] {$S_{0}$};
\path (A) edge [loop left] node[left] {$1-q_{0}\quad|\quad r=0$} (A);
\path (C) edge [bend left] node[below] {$q_{1}\quad|\quad r=0$} (A);
\path (A) edge [bend left] node[above] {$q_{0}\quad|\quad r=1$} (C);
\path (C) edge [loop right] node[right] {$1-q_{1}\quad|\quad r=1$} (C);
\end{tikzpicture}
\caption{The causal state diagram for a two-causal state process. The causal states, $S_0$ and $S_1$, are represented by a pair of circles. Arrows denote possible transitions. An arrow pointing from $S_j$ to $S_k$, $j,k \in \{0,1\}$, is labeled by the probability that the epsilon machine will transition from causal state $S_j$ to causal state $S_k$ upon emitting $x_t = r$:  $T^{r}_{j,k} $, we also explicitly specify the matching value of $r$. For example, the above process has $P(S_{0},\mathbf{X}_{t}=0|S_{1}) = T^{0}_{1,0} = q_1$ with accompanying emission $r= 0$.}\label{fig:markovchain}
\end{figure*}

 Other properties of statistical complexity are more puzzling. Using a simple example of a general two-state process depicted in Fig. \ref{fig:markovchain}, for $q_{0}=q_{1}= 0.5+\delta$, where $\delta > 0 $ is arbitrarily small, the output is very close to being completely random. Nevertheless, the conditional futures of the two causal states differ, and $C_{\mu}=1$. However, at $\delta=0$, $C_{\mu}$ is 0. Such discontinuity appears surprising for a measure of structure.


\subsection{Quantum is Simpler}\label{subsect:quantumstatcomplex}

All the above results assume our mathematical model processes classical information. Can a quantum extension of statistical complexity have different qualitative behavior?

Quantum logic allows the input information to be encoded into a quantum mechanical system. The advantage here is that this gives extra freedom to encode information in quantum superpositions. This lends a quantum refinement to the standard $\epsilon$-machine. In quantum $\epsilon$-machines each causal state is associated with a quantum causal state:
\begin{equation}\label{eq:quantumcausalstates}
\ket{S_{j}}=\sum_{k=1}^{|\mathcal{S}|}\sum_{r}\sqrt{T^{(r)}_{j,k}}\ket{r}\ket{k},
\end{equation}
where $\ket{r}$ belongs to a Hilbert space of dimension $|\Sigma|$ and $\ket{k}$ to a space of dimension $|\mathcal{S}|$. The quantum causal states, $\ket{S_{i}}$, are in general not orthogonal; nevertheless we can construct a systematic method to sample from $P(\righta{X}|\lefta{X} = \lefta{x})$, when given the appropriate quantum causal state $\ket{S_i = \epsilon(\lefta{x})}$.

To see this, consider a machine that takes  $\ket{S_i = \epsilon(\lefta{x})}$ directly as its input. It can generate the the correct output statistics at each time step,  by measuring $\ket{S_{j}}$ in the $\ket{r}$ basis. Each specific outcome $r$ occurs with probability $T_{j,k}^{(r)}$, and collapses the system to $\ket{k}$. The machine sets the output $x_{0}$ equal to $``r"$, and the quantum $\e$-machine then applies a quantum operation that maps $\ket{k}$ to state $\ket{S_{k}}$. Iterating this protocol will give a series of output values $x_{0}x_{1}x_{2}\cdots$ with the correct statistical distribution. This can be formalized as:

\begin{definition} The quantum $\e$-machine of a process $P(\lefta{X},\righta{X})$, is the ordered pair $\{\e_{q},\mathcal{S}_q\}$ where $\mathcal{S}_q = \{\ket{S_{i}}\}_{i=1,2,\dots, N}$ is the set of quantum causal states, and $\e_{q}$ is a function such that $\e_{q}(\cev{x})=\ket{S_{i}}$.
\end{definition}

The complexity of the resulting model is again determined by the entropy of its input that is given by the von Neumann entropy
\begin{equation}
C_{q}=-\Tr\rho\log\rho
\end{equation}
where $\rho=\sum_{i}p_{i}\ket{S_{i}}\bra{S_{i}}$ and $p_{i}$ is the probability of initiating the machine in state $\ket{S_{i}}$. Since the $\ket{S_{i}}$ are generally non-orthogonal, $C_q$ is often strictly less than $C_\mu$. Quantum $\e$-machines thus can have complexity below what is possible using any classical predictive model.

The advantage of quantum $\epsilon$-machines over their classical counterparts rests in the observation that future predictions do not require complete knowledge of the causal state the process started in. Indeed in the case of classical $\epsilon$-machines two instances of a process can start in different causal states, $S_i$ and $S_j$, where both $T^r_{i,k}$ and $T^r_{j,k}$ are non-zero. With probability $T^r_{i,k}T^r_{j,k}$ both  $\epsilon$-machines will transition to some coinciding causal state $S_k$ upon the same emission $r$. If this happens, all future statistics will be identical and no amount of future observations can ever fully identify whether the system started in $S_i$ or $S_j$. Thus some of the information used to distinguish between $S_i$ and $S_j$ is wasted.

Quantum mechanics allows the freedom to store different causal states as non-orthogonal quantum states, without employing classical randomness. Quantum $\epsilon$-machines exploit this -- they utilize quantum causal states that distinguish past causal states only to a degree sufficient for generating correct statistical behaviour. For example, for a process with two classical causal states $S_{0}$ and $S_{1}$, the emission alphabet $\Sigma = \{0,1\}$ and transition probabilities $T^{r}_{j,k} = \delta^r_k T_{jk}$ (where $\delta^r_k$ is the Kronecker delta) we have quantum causal states
\begin{eqnarray}\label{eqn:generalquantumstate}
\ket{S_{0}}&=\sqrt{T^{(0)}_{0,0}}\ket{00}+\sqrt{T^{(1)}_{0,1}}\ket{11}\\
\ket{S_{1}}&=\sqrt{T^{(0)}_{1,0}}\ket{00}+\sqrt{T^{(1)}_{1,1}}\ket{11}
\end{eqnarray}
Since we are dealing effectively with a two-dimensional space we label $\ket{00}$ as $\widetilde{\ket{0}}$ and $\ket{11}$ as $\widetilde{\ket{1}}$. If $S_1$ and $S_2$ have non-zero probability of transitioning to the same causal state, $S_k$, then clearly $T^{(k)}_{0,k}$ and $T^{(k)}_{1,k}$ are non-zero. Therefore $\braket{S_{1}}{S_{2}} > 0$, and hence $C_{q}\leq C_{\mu}$. Quantum $\e$-machines are simpler then their simplest classical alternatives, and this difference becomes ever more pronounced as the future statistics of their associated causal states becomes more similar.

\section{Complexity with Quantum Logic}\label{sect:quantcomplexity}

\subsection{Quantum $\e$-machines and Complexity}
How does the advent of quantum logic affect our original motivation of studying $\epsilon$-machines? The statistical complexity was proposed as a measure of structure in line with the ideal that the more memory required to model a process the greater its complexity. The above observations indicate that if we adopt such ideals we must necessarily accept that what we perceive to be complex depends on what information theory we use.

If we are limited to classical logic then the simplest way to model a process is through classical $\epsilon$-machines resulting in a perceived complexity of $C_\mu$. Thus, Crutchfield defined $C_\mu$ as the statistical complexity -- and motivated it as an intrinsic measure of structure and complexity of a given stochastic process $P(\lefta{X},\righta{X})$.

However, if we admit quantum information and quantum logic a stochastic process is likely to look simpler. If we are to model the process using a quantum $\e$-machine, then the perceived complexity of the system, $C_q$, is often strictly less than $C_\mu$. To a creature that employs quantum $\epsilon$-machines reality would appear simpler. Could $C_q$ provide an alternative quantifier of complexity?

This article explores the behavior of $C_q$. In studying how it evolves for a simple process we show how our notions of structure and complexity can diverge in the presence of quantum logic. We outline a scenario in which the behavior of $C_q$ aligns much more closely with our expectations of how  complexity  behaves.

We note, however, that while it is tempting to immediately name $C_q$ as the quantum statistical complexity, such an assignment would be rash. The question of whether quantum $\epsilon$-machines are the simplest quantum models remains open. In this article we refer to $C_q$ as the \emph{quantum $\e$-machine complexity} to avoid confusion, and leave questions of whether even simpler quantum models exist for future work.

\subsection{The Classical-Quantum Divergence}\label{subsect:limit}
To illustrate the divergence in the complexity of classical and quantum $\e$-machines, we outline a simple process involving a box containing a single coin. At each time-step $t$ the box is perturbed and the state of the coin is measured. This results in a binary output that is either heads ($x_t = 0$) or tails ($x_t = 1$), governed by random variable $\mathbf{X}_t$. The probability distribution over an infinite sequence of such measurements then defines a stochastic process $P(\lefta{X},\righta{X})$.

We assume that the act of perturbation may flip the state of the coin. The coin may be biased, such that perturbing a coin in state $k$ will cause it to flip with probability $0<q_k<1$, $k \in \{0,1\}$. Note that the special case of an unbiased coin was analyzed in \cite{gu2012quantum}, where it was referred to as the \emph{perturbed coin}.

The resulting process is clearly Markovian; all information about $\righta{X}$ is contained in $X_{-1}$. For $q_0,q_1\neq0.5$, we have two causal states. One equivalence class containing all pasts ending in ``0", signifying the last state was heads, and the other containing pasts ending in ``1". These causal states are denoted as $S_{0}$ and $S_{1}$. The transition probabilities for the corresponding $\epsilon$-machine are then given by $T^{(r)}_{jk} = \delta^r_k T_{jk}$, where $T_{jk}$ are elements of the matrix
\begin{equation} T=\begin{pmatrix}
1-q_{0}&q_{1}\\q_{0}&1-q_{1}\end{pmatrix}.
\end{equation}
Together these objects define the classical $\epsilon$-machine of the process (Fig~\ref{fig:markovchain}).

The statistical complexity of the process is then determined by $H(\mathbf{S}) = - p_0 \log_2 p_0 - p_1 \log_2 p_1$, where $p_i = P(\epsilon(\lefta{x}) \in S_i)$ for $i \in \{0,1\}$. To find these, let $\mathbf{p} = (p_0,p_1)$; then $\mathbf{p}$ satisfies $\mathbf{p}=T\mathbf{p}$, with solution
\begin{equation}\label{eq:steadystate1}
\mathbf{p}=\frac{1}{q_{1}+q_{0}}\left(\begin{array}{c}
q_{1}\\ q_{0}
\end{array}\right).
\end{equation}
Thus $p_{0}=\frac{q_{1}}{q_{0}+q_{1}}$ and $p_{1}=\frac{q_{0}}{q_{0}+q_{1}}$. The perturbed coin its own simplest model, with statistical complexity

\begin{equation}\label{eq:2statestatcomplex}
\begin{aligned}
C_{\mu}&=-\frac{q_{0}}{q_{0}+q_{1}}\log\left({\frac{q_{0}}{q_{0}+q_{1}}}\right)-\frac{q_{1}}{q_{0}+q_{1}}\log\left({\frac{q_{1}}{q_{0}+q_{1}}}\right).
		\end{aligned}
\end{equation}

\begin{figure}
\centering
\includegraphics[scale=0.35]{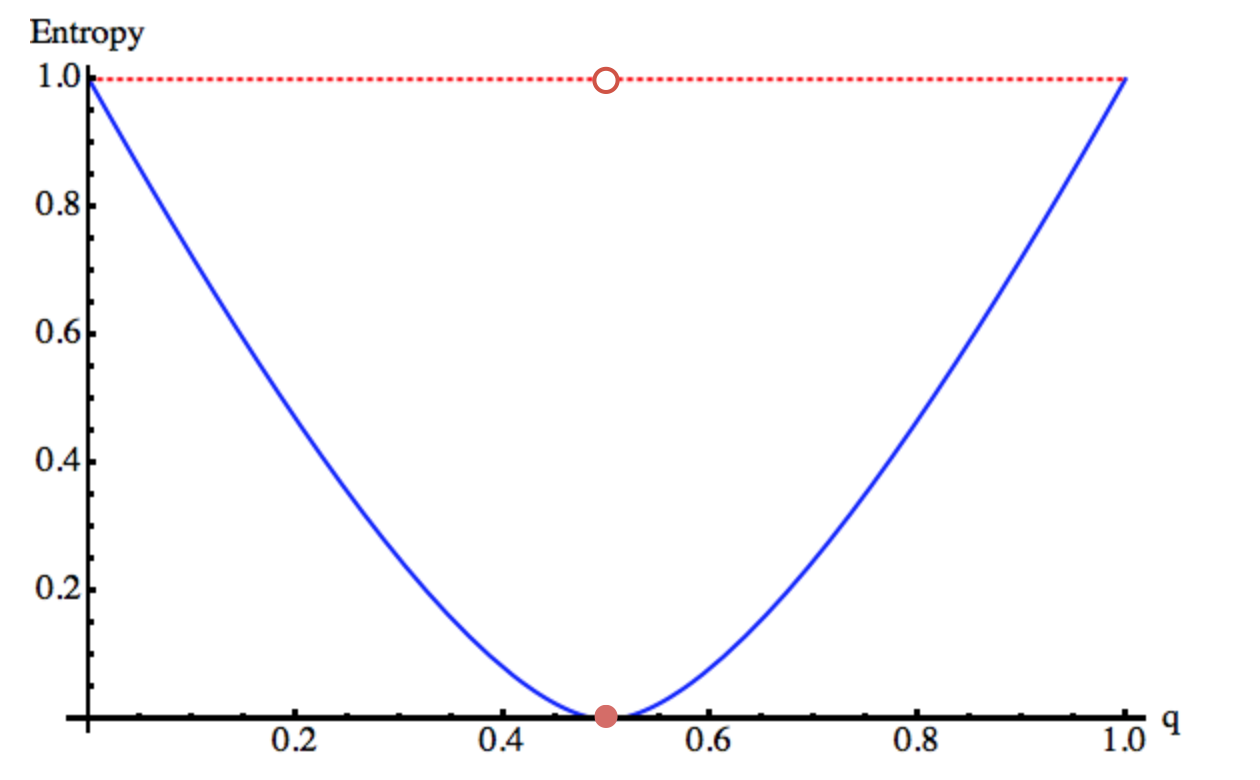}
\caption{ A plot of both the classical statistical complexity $C_{\mu}$ (red, dashed) and $C_q$ (blue, solid) for the perturbed coin, against the probability of flipping $q = q_1 = q_2$. At $q= 0.5$ the coin is completely fair and the output is random, making it unnecessary to store any information about the past to optimally predict the future. This plot illustrates the discontinuity in $C_{\mu}$  which jumps directly from 1 to 0 at q=0.5. In contrast $C_q$ is a continuous function of $q$.}\label{fig:perturbedcoin}
\end{figure}

By comparison the quantum causal states are $\ket{S_{0}}=\sqrt{1-q_0}\ket{0}+\sqrt{q_0}\ket{1}$ and $\ket{S_1}=\sqrt{q_1}\ket{0}+\sqrt{1-q_1}\ket{1}$. Thus the complexity of the quantum $\epsilon$-machine is $C_{q}=-\Tr\rho\log\rho$, with $\rho=p_0\ket{S_{0}}\bra{S_{0}}+p_1\ket{S_{1}}\bra{S_{1}}$.

The complexity of classical $\epsilon$-machines and their quantum counterpart are plotted in Fig~\ref{fig:perturbedcoin} for the special case where the coin is unbiased, i.e., $q_0 = q_1 = q$. The classical measure of complexity is clearly discontinuous: for $q \neq 0.5$, the two causal states are equiprobable and thus $C_\mu = 1$ \cite{crutchfield1997statistical}. Yet at $q = 0.5$, the process becomes completely random and thus the future is statistical identical for all pasts; there is only one causal state and $C_\mu = 0$. An infinitesimal perturbation in $q$ around $0.5$ leads to a sudden change in statistical complexity.

The complexity of the quantum $\epsilon$-machine, in contrast, remains continuous. As $q$ approaches $0.5$, the process becomes progressively more random; the overlap between the future statistics of the two causal states increase, thereby increasing the advantage of storing them in non-orthogonal states. In the limit $q \rightarrow 0.5$, $\braket{S_0}{S_1} \rightarrow 1$ and $C_q$ smoothly converges to $0$.

Thus, what appears in $\epsilon$-machines as a striking discontinuity in complexity vanishes when these machines are quantized. Conceptually, discontinuities in complexity appear difficult to explain. Why would an infinitesimal perturbation in the observed statistics leads to a large change in its perceived structure and complexity? The use of quantum $\epsilon$-machines seems to resolve this conundrum; and thus presents a promising refinement of their classical predecessors.

\section{The Thermalizing Qubit Cloud - a System of Ever Increasing Entropy}\label{sect:toymodel}
Can quantum $\epsilon$-machines be used to formalize our intuition of how complexity should evolve? We  shed light on this by studying how complexity evolves in a simple system featuring monotonically inreasing entropy  - a cloud of qubits undergoing gradual thermalization. Such an environment adheres to the second law of thermodynamics; mimicking our ideal that all systems graduate towards a state of total disorder.

Our intuition tells us that at either end of the scale complexity should be minimal. The system should evolve from something simple, to something more complex, and back. Should $C_q$ be a true quantifier of complexity, we would expect it to exhibit similar behavior.

Consider an environmental bath containing a large number of $(N>>1)$ qubits, with a global thermalization parameter $\lambda$. When $\lambda = 0$, the system is pure, and all qubits are in the state $\rho(0) = \ket{0}\bra{0}$. When $\lambda = 1$, all qubits are  maximally random with $\rho(1) = I/2$. With gradual thermalization, the system, monotonically evolves from order to disorder, such that each qubit  evolves according to
\begin{equation}\label{eq:qubit}
\rho_{e}(\lambda)=(1-\lambda)\ket{0}\bra{0}+\frac{\lambda}{2}I.
\end{equation}
The whole system can be described as $\rho_{e}^{\otimes N}(\lambda)$. Clearly this system exhibits monotonically increasing entropy, evolving smoothly from the minimum value of $0$ to the maximum value of $N$.
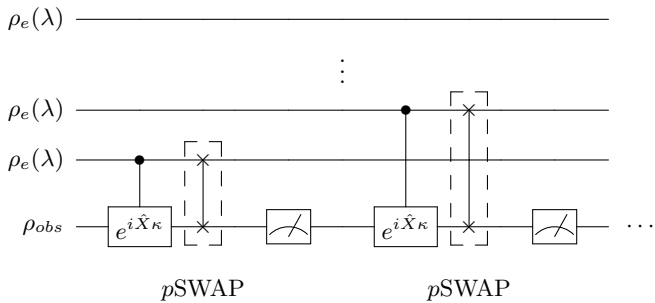
\begin{figure}[h!]
\begin{displaymath}
\Qcircuit @C=1.3em @R=1.8em {
& \lstick{\rho_{e}(\lambda)} & \qw & \qw &\qw &  \qw&  \qw  & \qw &\qw &\qw&\qw &\qw\\
& & &   & &  &{\vdots} &  & & & &\\
& \lstick{\rho_{e}(\lambda)} & \qw & \qw &\qw &  \qw&  \qw  & \ctrl{1} &\qswap &\qw&\qw &\qw\\
& \lstick{\rho_{e}(\lambda)} & \ctrl{1} & \qswap &\qw &  \qw&  \qw  & \qw &\qw\qwx &\qw&\qw&\qw \\
& \lstick{\rho_{obs}} &  \gate{{e^{i\hat{X}\kappa}}}   &  \qswap \qwx &\qw & \meter &\qw & \gate{e^{i\hat{X}\kappa}}  \qwx  &\qswap\qwx &\qw&\meter&\qw&\cdots \gategroup{4}{4}{5}{4}{1.5em}{--}
\gategroup{3}{9}{5}{9}{1.5em}{--}
\\
& & &  {p\mathrm{SWAP}} & &  & &  &{p\mathrm{SWAP}} & & &
} \end{displaymath}
\caption{Circuit representation of how the observer's qubit evolves through interactions with the qubit cloud. At each time step, $t$, a new ancillary environmental qubit $\rho_{e}(\lambda)$ interacts with the observer's probe $\rho_{obs}$. The interaction is modeled by a controlled unitary operation, $\ket{0}_e\bra{0}_e \otimes 1 + \ket{1}_e\bra{1}_e \otimes X\kappa $ where $\otimes $ is the direct product and $X\kappa = \exp{(i\hat{X}\kappa)}$ is a function of the Pauli $X$ operator, $\hat{X}$, and  an interaction strength parameter $0 \leq \kappa \leq \frac{\pi}{2}$. For the maximally entangling case $\kappa = \frac{\pi}{2}$ this interaction reduces to a controlled NOT gate. The pSWAP (probabilistic SWAP) defines a the execution of a SWAP gate, $U_s: \ket{\phi}\ket{\psi} \rightarrow \ket{\psi}\ket{\phi}$ with probability $g$; and models the observer's uncertainty of which qubit represents the probe after interaction. The symmetric case of $g = 0.5$ corresponds to the case where the observe completely loses track of which output qubit is which. At the end of the subroutine the observer measures his qubit in the $Z$ basis and outputs the answer as $x_t$. He then repeats the exercise with a new ancillary environmental qubit. }\label{fig:circuit}
\end{figure}
%

To characterize how complexity evolves in this system, we introduce an observer; an entity that probes this \systemname at particular values of $\lambda$. For this toy model we construct a scheme in which the observer's measurement device consists of a single probe qubit. This probe is engineered to interact with qubits within the cloud (specifics below), during which the observer monitors the probe at discrete time intervals to retrieve a sequence of binary outcomes $\cdots x_{-2}x_{-1}x_{0}x_{1}\cdots$. From these statistics the observe constructs a stochastic process $P(\lefta{X},\righta{X})$, which is then used to quantify the complexity of the environment.

The specifics of our measurement scheme are outlined in Fig. \ref{fig:circuit}. At each time-step $t$, the observer scatters his probe qubit $\rho_{obs}$ with a randomly selected qubit in the cloud; he then chooses one of the two output qubits with probability $0<g<1$ and measures it in the computational basis to generate output $x_{t}$. Repetition of this process then results in a bit-string whose distribution is governed by an associated stochastic process $P(\lefta{X},\righta{X})$. We make two simplifying assumptions about the system.\begin{enumerate}
\item The timescale  in which these measurements are made is infinitesimal compared to the timescale in which $\lambda$ evolves.
\item The number of qubits in the cloud, $N$, is sufficiently large that the probe quit never interacts with the same qubit twice.
\end{enumerate}
These assumptions ensure that that at each value of $\lambda$ we will probe the environment many times (collect many measurement outcomes) allowing the bit string we generate to be modeled by a stationary stochastic process. These assumptions apply, for example, in studying systems on the macroscopic scale.

Under these assumptions it is easy to see that the entropic properties of $P(\lefta{X},\righta{X})$ follow the entropic properties of the qubit cloud. At $\lambda = 0$, $P(\lefta{X},\righta{X})$ takes on unit probability for a sequence of $0$'s, and thus has zero entropy. Meanwhile, its entropy is maximal at $\lambda = 1$. Thus, the toy model provides a first order testing ground for our proposed measure of complexity.

\subsection{Statistical Complexity of the Qubit Cloud}\label{sect:models}
The first step in determining how complexity evolves in our \systemname is to construct its associated $\epsilon$-machine. To do this, we need to characterize the set of causal states $\mathcal{S}$ and associated transition probabilities. The first observation we make is that the process is Markovian. At each time step the equivalent quantum circuit is initialized in input state: $\ket{k}\bra{k}\otimes\rho_e(\lambda)$, where the last measurement outcome $x_{t-1} = k \in \{0,1\}$ completely determines the observer's input state $\ket{k}$. This input determines $\righta{X}$, for more detail see the Appendix. Hence all information about the  future $\righta{X}$, is contained in ${X}_{-1}$.

This Markovian property implies that the process has at most two causal states: $S_{0}=\{0,10,00,110,\cdots\}$, the set of pasts ending with ``0", and $S_{1}=\{1,01,11,011,\cdots\}$, the set of pasts ending with ``1". Indeed, provided the process is not completely random or completely uniform (i.e. $0<\lambda <1$), we must record something about the past to predict the future, i.e. $P(\righta{X}|S=S_{0}) \neq P(\righta{X}| S= S_{1})$. Thus the two causal states $S_{0}$ and $S_{1}$ are distinct.

To evaluate the transition probabilities $T^{(r)}_{jk}$, we first note that $T^{(r)}_{jk} = \delta^r_k T_{jk}$: if the system transits to $S_{k}$ at time $t$ it will always emit $x_t = k$. Thus the \systemname  generates a process $P(\lefta{X},\righta{X})$ that is statistically identical to the perturbed coin (see Fig.~\ref{fig:markovchain}) with $T_{jk}$ described by the transition matrix
\begin{equation} T=\begin{pmatrix}
1-q_{0}&q_{1}\\q_{0}&1-q_{1}\end{pmatrix},
\end{equation}
where $q_0$ and $q_1$ are $\lambda$-dependent. We evaluate these in the Appendix A to find
\begin{subequations}\label{eq:q1q2notation}
\begin{align}
q_{0}(\lambda)&=g\frac{\lambda}{2}+(1-g)\frac{\lambda}{2}\sin^{2}{(\kappa)},\\
q_{1}(\lambda)&=(1-g)\frac{\lambda}{2}   \sin ^{2}(\kappa)+g\left(1-\frac{\lambda }{2}\right).
\end{align}
\end{subequations}
The statistical complexity of the process is then defined by the complexity of this model; i.e., the entropy of the random variable $\mathbf{S}$ over causal states. Let
$p_i = P(\mathbf{S} = S_i)$ be the probability an $\epsilon$-machine is in state $S_i$, then from Eq. \eqref{eq:steadystate1};
\begin{equation}\label{eq:proportions}
 p_0(\lambda) = \frac{q_1(\lambda)}{q_0(\lambda) + q_1(\lambda)}, \qquad  p_1(\lambda) = \frac{q_0(\lambda)}{q_0(\lambda) + q_1(\lambda)},
\end{equation}
and the statistical complexity is thus
\begin{equation}\label{eq:cmachine}
C_\mu(\lambda) = -\sum_i p_i(\lambda) \log_2 p_i(\lambda), \qquad 0 < \lambda < 1.
\end{equation}
for the special case where $\lambda = 1$, $P(\lefta{X},\righta{X})$ becomes completely random and there is only a single causal state encompassing all possible pasts. Therefore the complexity of the system reduces to $0$, i.e., $C_\mu(1) = 0$.

\begin{figure}[htb]\begin{centering}
\includegraphics[scale=0.35]{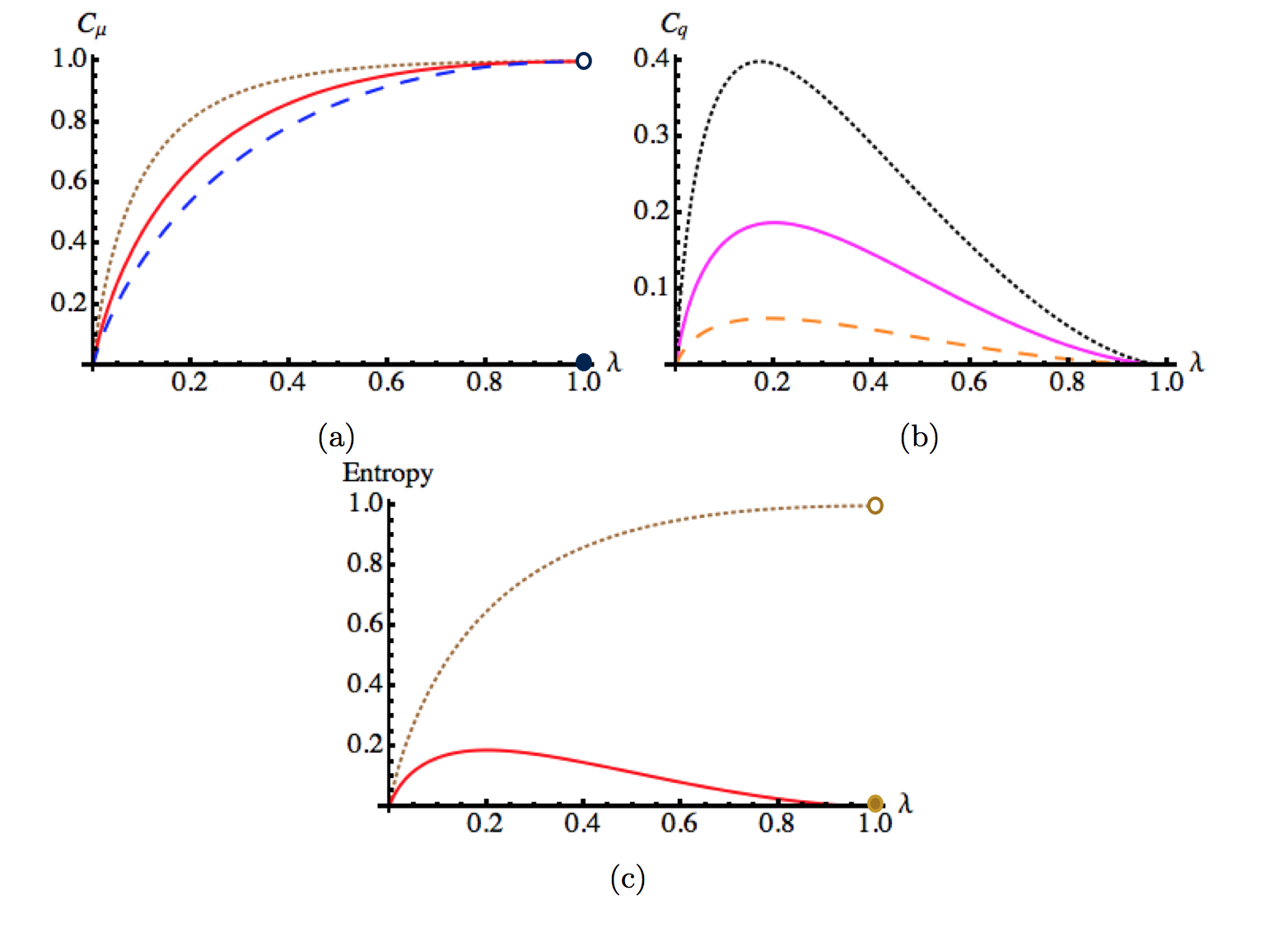}\caption{Overview of $C_q$ and $C_{\mu}$ for the thermalizing qubit cloud, as a function of the thermalization parrameter, $0<\lambda<1$. As $\lambda$  increases more noise is introduced and the environmental qubits, $\rho_e(\lambda)$, becomes more mixed. (a) Plot of $C_{\mu}$ vs. $\lambda$ for various swap probabilities, $g=0.25$ (brown, dotted), $g=0.5$ (red, solid) and $g=0.75$ (blue, dashed). Qualitatively $C_{\mu}$ is a monotonically increasing function of $\lambda$. (b) Plot of $C_{q}$ vs. $\lambda$ for values of $g=0.25$ (black, dotted), $g=0.5$ (purple, solid) and $g=0.75$ (orange, dashed). $C_{q}$ dies down as $\lambda \rightarrow 1$ and the corresponding stochastic process becomes more random. (c) Plot of $g=0.5$ case for both  $C_{\mu}$ (brown, dotted) and $C_q$ (red, solid)   to illustrate the comparative behaviour.}\label{fig:cmucq}
\end{centering}\end{figure}

\subsection{Quantum $\e$-machines and their Complexity}\label{sect:models}

The quantum $\epsilon$-machine can be determined by directly quantizing the causal states; resulting in two quantum causal states
\begin{subequations}
\begin{align}
\ket{S_{0}}&=\sqrt{1-q_{0}}\widetilde{\ket{0}}+\sqrt{q_{0}}\widetilde{\ket{1}},\\
\ket{S_{1}}&=\sqrt{q_{1}}\widetilde{\ket{0}}+\sqrt{1-q_{1}}\widetilde{\ket{1}}.
\end{align}
\end{subequations}
The complexity of the quantum $\epsilon$-machine is given by the von Neumann entropy of the density operator
\begin{equation}
\rho=p_{0}\ket{S_{0}}\bra{S_{0}}+p_{1}\ket{S_{1}}\bra{S_{1}},
\end{equation}
where $p_{i}$ is the probability of finding the quantum $\e$-machine in the state $\ket{S_{i}}$. This probability satisfies Eq. \eqref{eq:proportions}.

\subsection{Complexity Dynamics}\label{subsubsect:cnot}
We now have all the tools necessary to analyze how complexity evolves within the qubit cloud--both through the lens of conventional $\epsilon$-machines and their quantum mechanical counterpart.
\begin{figure*}[htb!]
\begin{centering}
\includegraphics[width=0.8\textwidth]{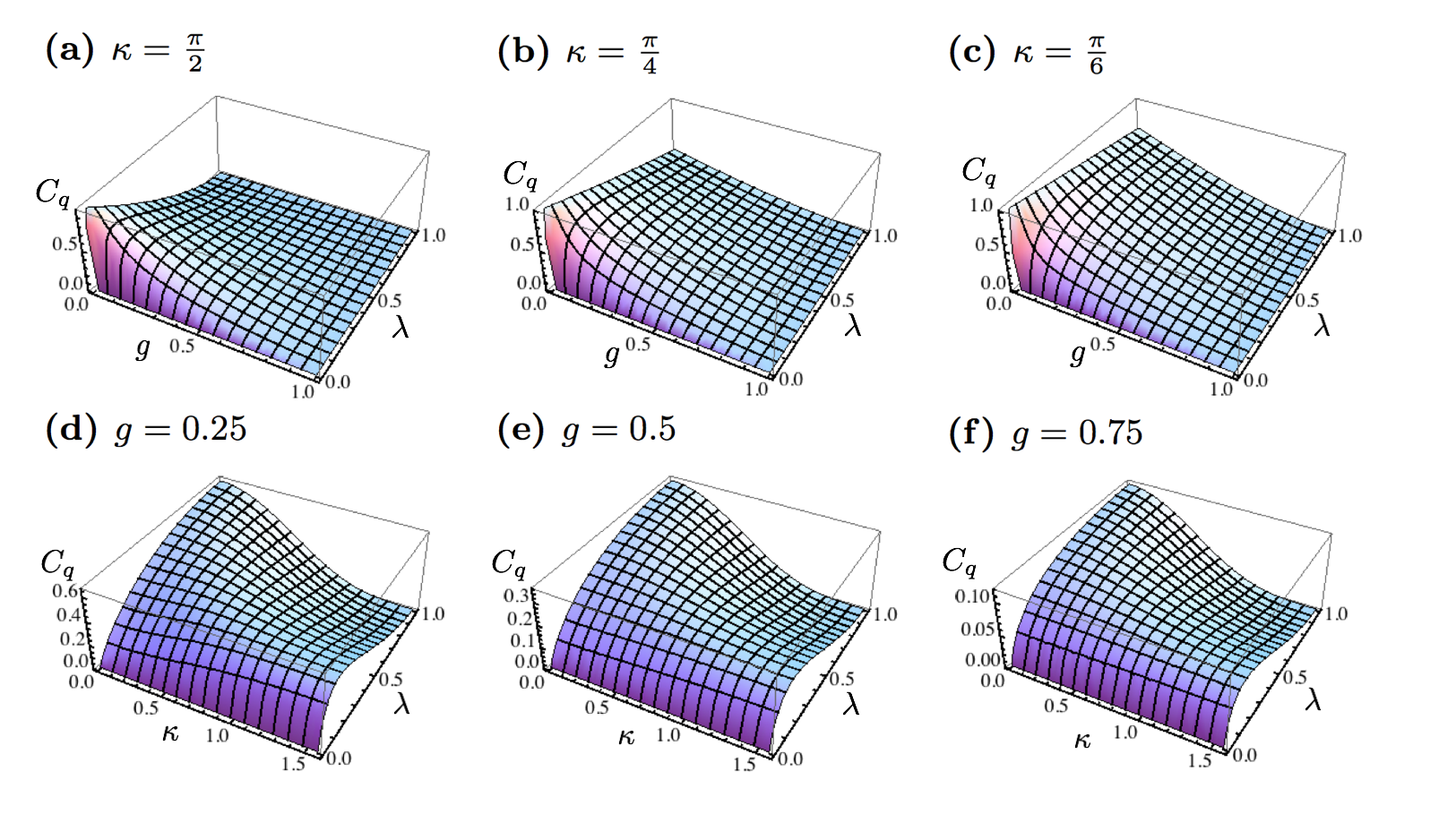}
\caption{Evolution of $C_q$ with thermalization parameter $\lambda$ for various values of interaction strength $\kappa$ and swap probability $g$. In (a), (b), (c), $g$ is varied for various fixed values of $\kappa$. In (d), (e), (f), $\kappa$ is varied for various fixed values of $g$.}\label{fig:generalcnot}
\end{centering}\end{figure*}

Firstly, observe that the complexity of the system, regardless of which measure we use ($C_\mu$ or $C_q$), is indeed minimal for $\lambda = 0,1$. Eq. \eqref{eq:q1q2notation} implies that when $\lambda = 0$, $q_0(0) = 0$ and $q_1(0) > 0$; thus the $\epsilon$-machine resides in state $S_0$ with unit probability. The process is trivial (the output is a uniform string of $0$s), resulting in $C_\mu = C_q = 0$. On the other end, $q_0 = q_1$ when $\lambda = 1$, thus $P(\righta{X} | \mathbf{S} = S_0) = P(\righta{X} | \mathbf{S} = S_1)$. This indicates that the two causal states have statistically identical futures; and therefore collapse into a single causal state that encompasses all pasts. Again $C_\mu = C_q = 0$. Thus, both complexity measures agree with our intuition that a system at either $0$ or maximal entropy process is completely trivial.

The behavior of $C_q$ and $C_\mu$, however, diverges for intermediate values of $\lambda$. Fig. \ref{fig:cmucq} displays the generic behavior of both $C_\mu$ and $C_q$. $C_\mu$ is a monotonically increasing function for all $\lambda \neq 1$. In fact, $\lambda = 1$ is a point of discontinuity; where $C_\mu$ drops sharply to $0$. It indicates that the process is `maximally complex' an infinitesimal distance away from being completely random. If we put this in the context of a physical system where complete randomness is an idealized limit, it seems to suggest that complexity shares almost identical behavior with entropy. These results contrast quite sharply with our perceived conception of complexity, and intuition that complexity should be complementary to randomness \cite{crutchfield2012between,crutchfield1994calculi}. The behavior of $C_q$, on the other hand, appears in line with what we expect. The quantum $\e$-machine complexity rises and falls continuously, taking on a peak value at $\lambda \approx 0.2$. There is no sudden jumps, and the complexity of the qubit cloud near complete thermalization is indeed very close to $0$.

Fig. \ref{fig:cmucq} (c) compares $C_q$ and $C_\mu$ directly for the special case where $g = \frac{1}{2}$ and $\kappa = \pi/2$. This corresponds to the limiting scenario where the probe interacts with the qubit cloud through the idealized CNOT gate and the observer measures one of the two outputs at random. It is clear that the quantum $\epsilon$-machine is able to model the resulting process using far less memory; and this advantage grows with $\lambda$. These results are not isolated to specific parameter choices. Fig. \ref{fig:generalcnot} displays the behavior of $C_q$ for various values of $\kappa$ and $g$. In general, $C_{q}$ is a continuous function of $\lambda$ and gradually diminishes as $\lambda$ approaches either $0$ or $1$, while peaking somewhere in the middle.

These results can be understood qualitatively. As the process becomes more random there are two contributing sources to the complexity. The first is positive: noise gives the system the energy to transition from causal state $S_0$ to $S_1$. This applies an equalizing pressure to the population level of two causal states $S_0$, and $S_1$. Thus, more memory is require to distinguish the two possibilities. The second a negative contribution. As noise increases; the future statistics of $S_0$ and $S_1$ become less distinct, and thus distinguishing between the two is less meaningful. The evolution of complexity within the system rests on the balance of these two contributions.

The classical statistical complexity, however, accounts only for the former contribution. It is based entirely on the entropy of distinguishing causal states $S_0$ and $S_1$. Within our toy model as $\lambda$ increases $p_0 \rightarrow p_1$. Thus the statistical complexity undergoes monotonic gain. The progressive convergence in the conditional future of these two causal states remains ignored until $\lambda = 1$. $C_q$, on the other hand, continually accounts for the latter contribution. As the conditional futures of $S_0$ and $S_1$ converge, so do the quantum causal states $\ket{S_0}$ and $\ket{S_1}$. Thus the amount of memory required to model the process evolves continuously reflecting our ideal that complexity rests on the balance between order and disorder.

\section{Discussion}\label{sect:discussion}
In this article we explored a quantifier of complexity by extending the framework of $\epsilon$-machines into the quantum mechanical regime. In computational mechanics the information content of $\epsilon$-machines presents a popular approach to quantifying the structure of a given stochastic process, the rationale being that they are its simplest models. The advent of quantum $\epsilon$-machines, however, demonstrated that simpler models do exist. This motivated us to ask ``how complex would a stochastic process look to a quantum $\epsilon$-machine?"

Our results demonstrated a marked divergence in the complexity of $\epsilon$-machines and their quantum mechanical counterparts. In a process at various different stages of thermalization $\lambda$ we found that the statistical complexity increased monotonically with $\lambda$. Only at infinite temperature did it drop discontinuously to zero. The quantum $\epsilon$-machine complexity, on the other hand, behaved as a smooth function of $\lambda$ first rising for low values of entropy, then falling as the system approached total randomness.

When we envision the dynamics of complexity the latter quantity seems more reasonable. If we are to estimate the probability distribution of a given stochastic process $P(\lefta{X},\righta{X})$ through observations, there will always be a statistical margin of error. Thus, it is unsatisfactory for a quantity that describes the process complexity to jump to a fixed value (here $1$) for an infinitesimal perturbations away from total randomness. The continuity in the complexity of quantum $\epsilon$-machines is thus a welcome trait.

This article only skims the surface of how quantum theory may combine with computational mechanics. Many open questions remain. On the one hand, the model that we have presented here is but a toy with very specific assumptions on how the qubit cloud was probed. Could similar techniques be applied to more complex systems that have been studied within the framework of computational mechanics? We note that previous studies of statistical complexity in the Ising lattice demonstrated similar behavior~\cite{crutchfield1997statistical}. The complexity rose monotonically with temperature and only dropped to $0$ via a sudden discontinuous jump at infinite temperature. Could quantizing $\epsilon$-machines remove this discontinuity?

On the other hand there is currently no proof that quantum $\epsilon$-machines are the provably simplest quantum models. If this turns out false, would the true minimal amount of memory, $C_Q$ required to simulate a given stochastic process share the qualitative features of $C_q$? Certainly, using bounding arguments (As $0 \leq C_Q \leq C_q$; we can see that $C_Q$ has to be continuous at the point of maximal randomness in the qubit cloud.

The ultimate goal would be to present an operationally meaningful, yet nevertheless computable, quantifier of complexity. This would substantiate our intuition that complexity lies at the border between order and chaos - and thus pave the path for developing universal laws that govern complexity. This article presents one clue in the big puzzle - our notions of what is complex is affected by what sort of information we use; and quantum information could be a valuable tool in understanding what around us is ultimately complex.

{\it Acknowledgements.}--- The authors acknowledges helpful discussions with Karoline Weisner, Alex Monras and Borivoje Daki\'c. This work is supported in part by \chinagrant, the Singapore Ministry of Education and the Academic Research Fund Tier 3 MOE2012-T3-1-009 ``Random numbers from quantum processes".

\appendix
\section*{Appendix}

\subsection*{A. Evaluation of Perturbation Parameters for the \systemname}

In this Appendix we give a more in depth treatment of the \systemname.  The circuit representation of this model can be broken down into two stages: a two-qubit interaction between the observer's qubit and the environmental qubit, followed by a probabilistic swap defined in the caption of Fig.~\ref{fig:circuit}. Formally during the first stage the two-qubit unitary is given by
\begin{equation}
C_{X\kappa} = 1\otimes \ket{0}_e\bra{0}_e +X\kappa \otimes \ket{1}_e\bra{1}_e,
\end{equation}
 where $X\kappa = \exp{(i\hat{X} \kappa)}$ is defined in terms of the Pauli $X$ operator, and  $\otimes$ denotes the direct product.

The probabilistic SWAP operation acts on the combined system consisting of the environmental qubit and the observer's qubit, which we denote by $\rho_{obs,e}$. We define this transformation through
\begin{equation}\label{eq:pswap}
\rho_{obs,e}\rightarrow \rho_{obs,e}'=g\mathrm{U}_{\text{S}}\rho_{obs,e}\mathrm{U}_{\text{S}}^{\dagger}+(1-g)\rho_{obs,e}
\end{equation}
where $\mathrm{U}_{S}$ is the standard unitary SWAP operation defined by $U_{\text{S}} \ket{\phi}_{obs}\ket{\psi}_e = \ket{\psi}_{obs}\ket{\phi}_e$, and $g$ parametrises the  probability of swapping.

The assumption that the bath is extremely large, such that the observer's qubit never interacts with the same environmental qubit twice, implies that the system is initialized in a product state $\rho_{obs}\otimes\rho_{e}(\lambda)$. At each time step, $t$, the observers qubit is initialized in the quantum state corresponding to the last measurement outcome $x_{t-1}$, while the environmental qubit is given by \eqref{eq:qubit}.

Explicitly, if the outcome of the last measurement is $x_{t-1}=k$ (for $k \in \{0,1\}$), then the observer's qubit is initialized in state $\ket{k}$ and the combined two-qubit system, $\rho_{obs,e}$, is initialized as
\begin{align}\label{eq:inputstate0}
\rho_{obs}\otimes\rho_{e}(\lambda)&=\ket{k}\bra{k}\otimes\left[(1-\lambda)\ket{0}\bra{0}+\frac{\lambda}{2}\mathbb{1}\right]\\
&=(1-\frac{\lambda}{2})\ket{k0}\bra{k0}+\frac{\lambda}{2}\ket{k1}\bra{k1}.
\end{align}
 If $x_{t-1} = 0$ then after going through the $C_{X\kappa}$ interaction and the probabilistic SWAP operation we take the partial trace over the environmental qubit to recover the state of observer's qubit directly before measurement:
\begin{eqnarray}\label{eq:updatethermalcloud0}
&&\rho'_{obs}=\Tr_{e}(\rho'_{e,obs}) =\left[\left(1-\frac{\lambda}{2}\right)+\frac{\lambda}{2}\cos^{2}\kappa\right]\ket{0}\bra{0} +\notag\\
&& \left[g\frac{\lambda}{2}+ (1-g)\frac{\lambda}{2}\sin^{2}\kappa\right]\ket{1}\bra{1}+ i\sin \kappa\cos \kappa(\ket{1}\bra{0}-\ket{0}\bra{1}).\notag\\
\end{eqnarray}
Correspondingly, if $x_{t-1} = 1$ then
\begin{eqnarray}\label{eq:updatethermalcloud1}
\rho'_{obs}&=&\Tr_{e}(\rho'_{obs}\otimes\rho'_{e})\notag\\
&=&\left[g\left(1-\frac{\lambda}{2}\right)+(1-g)\frac{\lambda}{2}\sin^{2}\kappa\right]\ket{0}\bra{0}+\notag\\ &&\left[g\frac{\lambda}{2}+(1-g)\left(1-\frac{\lambda}{2}\sin^{2}\kappa\right)\right]\ket{1}\bra{1}+\notag\\&&(1-g)i\sin \kappa\cos \kappa(\ket{0}\bra{1}-\ket{1}\bra{0}).
\end{eqnarray}
The terms before $\ket{0}\bra{0}$ and $\ket{1}\bra{1}$ give the probabilities of measuring $\ket{0}$ and $\ket{1}$ and hence the statistics of the next output. The Markovian nature of the $\epsilon$-machine is clearly demonstrated by the fact that the probability $x_t= 0$ (or $1$) depends only on the value of $x_{t-1}$, the last measurement outcome. This establishes the set of causal states as  $S_{0}=\{0,10,00,110,\cdots\}$, the set of pasts ending with ``0", and $S_{1}=\{1,01,11,011,\cdots\}$, the set of pasts ending with ``1". From these results we can also find the transition probabilities for the corresponding stochastic process.

\subsection*{B. The Special Case of Maximally Interacting Probes}\label{subsubsect:cnot}
\begin{figure}[h]
\centering
\begin{tikzpicture}[-latex ,auto ,node distance = 4cm and 5cm ,on grid ,
semithick ,
state/.style ={ circle ,top color =white ,
draw, minimum width =1 cm}]
\node[state] (C) {$S_{1}$};
\node[state] (A) [left=of C] {$S_{0}$};
\path (A) edge [loop left] node[below = 0.3 cm] {$1-\frac{\lambda}{2}\, |\, r=0$} (A);
\path (C) edge [bend left] node[below =0.1 cm] {$g(1-\frac{\lambda}{2})+(1-g)\frac{\lambda}{2}\, |\, r=0$} (A);
\path (A) edge [bend left] node[above] {$\frac{\lambda}{2}\, |\, r=1$} (C);
\path (C) edge [loop right] node[left = 1.5 cm] {$g\frac{\lambda}{2}+(1-g)(1-\frac{\lambda}{2})\,|\, r=1$} (C);
\end{tikzpicture}
\caption{Causal state diagram for the thermalizing qubit cloud with interaction strength set by $\kappa=\frac{\pi}{2}$. The two causal states are denoted $S_0$ and $S_1$ and an arrow from $S_j$ to $S_k$ represents the corresponding transition, with label denoting the transition probability $T^r_{jk}$ and corresponding emission $r$.  }\label{fig:cnotemach}
\end{figure}
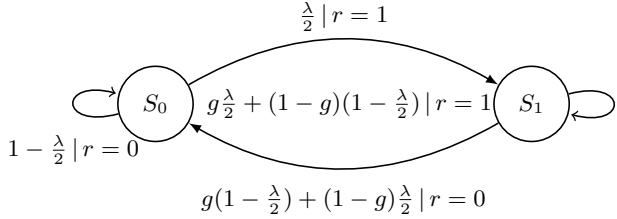

It is instructive to first outline special case where where $\kappa=\pi/2$.  Using the value of $\kappa = \pi/2$ in Eq. \eqref{eq:updatethermalcloud0} and \eqref{eq:updatethermalcloud1} we can simplify the transition probabilities to:

\begin{subequations}\label{eq:cnottransprob}
\begin{align}
P(\mathbf{S}_{t}=S_{0}|\mathbf{S}_{t-1}=S_{0})&=1-\frac{\lambda}{2}, \label{eq:p00}\\
P(\mathbf{S}_{t}=S_{1}|\mathbf{S}_{t-1}=S_{0})&=\frac{\lambda}{2},\label{eq:p10}\\
P(\mathbf{S}_{t}=S_{0}|\mathbf{S}_{t-1}=S_{1})&=g\left(1-\frac{\lambda}{2}\right)+(1-g)\frac{\lambda}{2},\label{eq:p01}\\
P(\mathbf{S}_{t}=S_{1}|\mathbf{S}_{t-1}=S_{1})&=g\frac{\lambda}{2}+(1-g)\left(1-\frac{\lambda}{2}\right).\label{eq:p11}
\end{align}
\end{subequations}
The $\e$-machine of the process is presented in Fig.~\ref{fig:cnotemach}.
The quantum causal states are subsequently reduced to
\begin{eqnarray}\label{eq:qcausalstates}
\ket{S_{0}}&=&\sqrt{1-\frac{\lambda}{2}}\widetilde{\ket{0}}+\sqrt{\frac{\lambda}{2}}\widetilde{\ket{1}},\\
\ket{S_{1}}&=&\sqrt{g(1-\frac{\lambda}{2})+(1-g)\frac{\lambda}{2}}\widetilde{\ket{0}}+\notag\\&&\sqrt{g(\frac{\lambda}{2})+(1-g)(1-\frac{\lambda}{2})}]\widetilde{\ket{1}}.
\end{eqnarray}
The resulting state of the quantum $\epsilon$-machine is:
\begin{equation}\label{eq:toydensitymatrix}
\rho=p_{0}\ket{S_{0}}\bra{S_{0}}+p_{1}\ket{S_{1}}\bra{S_{1}}.
\end{equation}
Substituting $q_{0}=\frac{\lambda}{2}$ and $q_{1}=g(1-\frac{\lambda}{2})+(1-g)\frac{\lambda}{2}$ into Eq.~\eqref{eq:steadystate1} directly yields
\begin{subequations}
\begin{align}
p_{0}&=\frac{-2g(\lambda-1)+\lambda}{2(g+\lambda-g\lambda)},\\
p_{1}&=\frac{\lambda}{2(g+\lambda-g\lambda)}.
\end{align}
\end{subequations}

\end{document}